\documentclass[aps,pra,preprint,superscriptaddress,showpacs]{revtex4}

\usepackage{graphicx}
\usepackage{bm} 
\usepackage{amsmath}
\usepackage{amssymb}

\newcommand{\half}{\frac{1}{2}}
\newcommand{\bOmega}{\bar{\Omega}}

\def\jcp#1#2#3{J.~Chem.~Phys.~{\bf #1},\ #2\ (#3)}

\def\pra#1#2#3{Phys.~Rev.~A~{\bf #1},\ #2\ (#3)}
\def\prl#1#2#3{Phys.~Rev.~Lett.~{\bf #1},\ #2\ (#3)}

\newcommand{\threejm}[6]{ \left(\begin{array}{ccc} #1 & #3 & #5\\
                                              #2 & #4 & #6
                                \end{array}
                          \right)}

\begin{document}

\title{Dynamics of OH($^2\Pi$)--He collisions in combined electric and magnetic fields}

\author{Timur V. Tscherbul}
\email{tshcherb@cfa.harvard.edu}
\affiliation{Institute for Theoretical Atomic, Molecular, and Optical Physics, Harvard-Smithsonian Center for Astrophysics, Cambridge, Massachusetts, 02138, USA}
\affiliation{Harvard-MIT Center for Ultracold Atoms, Cambridge,
  Massachusetts 02138, USA}
\author{Gerrit C. Groenenboom}
\email{gerritg@theochem.ru.nl}
\affiliation{Theoretical Chemistry, Institute for Molecules and Materials, Radboud University Nijmegen, Heyendaalseweg 135, 6525 AJ Nijmegen,
Netherlands}
\author{Roman V. Krems}
\affiliation{Department of Chemistry, University of British Columbia,
  Vancouver, British Columbia V6T 1Z1, Canada}
\author{Alex Dalgarno}
\affiliation{Institute for Theoretical Atomic, Molecular, and Optical Physics, Harvard-Smithsonian Center for Astrophysics, Cambridge, Massachusetts, 02138, USA}
\affiliation{Harvard-MIT Center for Ultracold Atoms, Cambridge,
  Massachusetts 02138, USA}

\date{\today}

\begin{abstract}
We use accurate quantum mechanical calculations to analyze the effects of parallel electric and magnetic fields on collision dynamics of OH($^2\Pi$) molecules.
It is demonstrated that spin relaxation in $^3$He--OH collisions at temperatures below 0.01 K can be effectively suppressed
by moderate electric fields of order 10 kV/cm. We show that electric fields can be used to manipulate Feshbach resonances in collisions of cold molecules.
Our results can be verified in experiments with OH molecules in Stark decelerated molecular beams and electromagnetic traps. 

\end{abstract}

\pacs{33.20.-t, 33.80.Ps}

\maketitle

\section{Introduction}

The rapid progress in the research field of cold molecules holds great promise for new and important discoveries at the interface of physics
and chemistry \cite{Zoller,Hinds,OHspectroscopy,Hudson,XeOH,Sawyer,RomanReview}.
The remarkable properties of cold molecules allow for their application in quantum information
processing, condensed-matter physics, physical chemistry, and precision spectroscopy. While some of these applications 
rely on specific properties of molecular ensembles, others exploit the diversity of molecular electronic states and energy level structures.
For example, recently proposed schemes for quantum computing with polar $^2\Sigma$ molecules in optical lattices make use of intermolecular
dipole-dipole interactions and molecule-field
couplings to enable communication between quantum bits \cite{Zoller}. The experiments aiming to measure the electric dipole moment of the electron
based on the spectroscopic study of cold YbF radicals exploit the relativistic distortion of molecular orbitals \cite{Hinds}.
The fine and hyperfine perturbations in the spectra of $^2\Pi$ molecules allow for the development of new frequency standards
and tests of physics beyond the Standard Model \cite{OHspectroscopy,Hudson}.

Particularly noteworthy are the experiments probing
inelastic energy transfer and chemical reactions of cold molecules \cite{XeOH,Sawyer,RomanReview}. The measurements of molecular
collision properties at low temperatures are important for the development of buffer gas cooling techniques, which rely on
collisional thermalization of molecules in a cell filled with cryogenic $^3$He gas. This technique produces molecules at
temperatures between 0.1 K and 0.5 K and allows for further evaporative cooling after removal of the buffer gas.
In order to permit observations and evaporative cooling, molecules must remain trapped for a long time, which is possible
if the ratio of elastic to inelastic spin-changing collisions exceeds $10^4$ \cite{Ketterle}. Previous experimental \cite{CaH,NH}
and theoretical \cite{Roman2004} work has shown that this requirement is generally satisfied for $\Sigma$-state diatomic molecules
with large rotational constants. However, it is unclear whether molecules in electronic states other than $\Sigma$ may be 
sympathetically cooled in a buffer gas cell. The mechanism of spin relaxation in collisions of molecules in $^2\Pi$
electronic state remains unknown.

The OH radical is one of the first molecular species that were cooled and trapped at milliKelvin temperatures
\cite{XeOH,Sawyer,OH1,GerritRadiativeLifetime,blackbody}.
The ground electronic state of OH is of $^2\Pi$ symmetry, and the energy levels of the molecule depend linearly
on the strength of an applied electric field above $\sim$5 kV/cm.
As a result, the OH radicals can be efficiently decelerated,
trapped, and manipulated using moderate static or time-varying electric fields available in the laboratory \cite{XeOH,OH1,GerritRadiativeLifetime,blackbody}.
A variety of novel experiments with trapped OH molecules has been
reported. The ability to fine tune the collision energy of a Stark-decelerated beam allowed Meijer and co-workers to study threshold
behavior of Xe-OH rotationally inelastic scattering at collision energies as low as 50 cm$^{-1}$ \cite{XeOH}. High-precision spectroscopy
of cold trapped OH was used to study the time evolution of the fine structure constant \cite{Hudson}, measure the lifetimes of vibrationally
excited states \cite{GerritRadiativeLifetime}, and determine the rates of optical pumping due to blackbody radiation \cite{blackbody}.
Sawyer {\it et al.} have recently reported measurements of cross sections for elastic and inelastic collisions of magnetically trapped OH molecules
with He atoms and D$_2$ molecules at kinetic energies of 60 cm$^{-1}$ and above \cite{Sawyer}.

Theoretical studies of low-energy collisions of OH molecules have been reported by several groups \cite{XeOH,Bohn1,Bohn2,Bohn3,Bohn4,Gianturco}.
Avdeenkov and Bohn studied ultracold collisions of OH molecules \cite{Bohn1,Bohn2} and discovered weakly-bound dimer states supported
by the dipole-dipole interaction forces in the presence of an external electric field \cite{Bohn2}. Ticknor and Bohn found a significant suppression of
inelastic relaxation rates for the low-field-seeking states of OH in a magnetic field \cite{Bohn3}. Lara {\it et al.} analyzed the
effects of non-adiabatic and hyperfine effects on field-free Rb--OH collisions \cite{Bohn4}. Their results
indicated that sympathetic cooling of OH molecules by collisions with Rb atoms might be challenging due to large inelastic loss rates.
Gonz{\'a}lez-S{\'a}nchez, Bodo, and Gianturco \cite{Gianturco} considered field-free collisions of rotationally excited OH molecules with He atoms and found
sharp propensity rules for rotational and $\Lambda$-doublet changing transitions at ultracold temperatures.

Here, we present a theoretical analysis of OH($^2\Pi$) collision dynamics in {\it combined} electric and magnetic fields.
We have previously demonstrated that spin-changing collisions of $\Sigma$-state molecules can be efficiently manipulated by
superimposed electric and magnetic fields \cite{OurWork1,OurWork2,JCP}. Building on our previous work \cite{OurWork1,OurWork2}
and the results of Bohn and co-workers \cite{Bohn1,Bohn2,Bohn3}, we develop a rigorous quantum theory of collisions between
$^2\Pi$ molecules and structureless atoms in external fields and calculate the dependence of the cross sections for He--OH collisions
on electric and magnetic fields. Our results suggest that collisions of OH molecules with He atoms can be efficiently
manipulated with the external fields. In particular, we demonstrate an efficient mechanism for suppression
of spin relaxation in $^2\Pi$ molecules with electric fields.


\section{Theory}

The quantum mechanical formalism for collisions of diatomic molecules in $^2\Pi$ electronic states in the absence of external
fields has been presented by several authors (see, e.g., Refs. \cite{Gianturco,MillardJCP,MillardCP}). Here, we focus on the theoretical aspects
relevant for incorporating the effects of electromagnetic fields in scattering calculations. Section IIA presents the discussion of
the influence of the electric and magnetic fields on the energy level structure of $^2\Pi$ molecules. In Sec. IIB, we
discuss the Hamiltonian of the collision complex and the coupled-channel representation of the scattering wave function.
A derivation of the matrix elements for the interaction potential operator in the basis of scattering states is presented in Sec. IIC.

\subsection{The OH molecule in superimposed electric and magnetic fields}

The Hamiltonian for a $^2\Pi$ molecule such as OH can be written as \cite{Brown1,Brown2,Gerrit}
\begin{equation}\label{Hmol}
\hat{H}_\text{mol} = \hat{H}_\text{rot} + \hat{H}_\text{SO} + \hat{H}_\Lambda + \hat{H}_\text{E} + \hat{H}_\text{B},
\end{equation} 
where $\hat{H}_\text{rot}$ is the angular part of the rotational kinetic energy \cite{GerritO2}
\begin{equation}\label{Hrot}
\hat{H}_\text{rot}=B_e(\hat{J}^\text{SF} - \hat{L}^\text{SF} - \hat{S}^\text{SF})^2,
\end{equation} 
$B_e$ is the rotational constant, $\hat{J}^\text{SF} = \hat{N}^\text{SF} + \hat{L}^\text{SF} + \hat{S}^\text{SF}$ is the total angular momentum,
$\hat{N}^\text{SF}$ is the rotational angular momentum
of the nuclei, $\hat{L}^\text{SF}$ is the electronic orbital angular momentum, and $\hat{S}^\text{SF}$ is the electron spin.
In Eq. (\ref{Hmol}), we have neglected the hyperfine interaction due to the nuclear spin of H. The hyperfine interaction constant
of OH is an order of magnitude smaller than the $\Lambda$-doublet splitting, and the hyperfine effects may alter collision
dynamics at temperatures below 4 mK \cite{Bohn3}. 
The angular momentum operators in Eq. (\ref{Hrot}) are defined in the space-fixed frame. However,
the symmetry properties of the electronic wave functions are most conveniently exploited in the molecule-fixed frame, with
the $z$-axis oriented along the OH bond. The row vector of molecule-fixed angular momentum operators can be defined as
$\hat{J}^\text{SF}=\hat{J}^\text{MF}\mathsf{R}(\bar{\alpha},\bar{\beta},0)$, where $\mathsf{R}$ is the matrix of direction cosines 
and $(\bar{\alpha},\bar{\beta})$ are the Euler angles which specify the orientation of the diatomic molecule in the space-fixed coordinate system.
Alternatively, one can define the column vector $\hat{J}^\text{SF}=\mathsf{R}(\bar{\alpha},\bar{\beta},0)\hat{J}^\text{MF}$
\cite{GerritO2}. The molecule-fixed angular momentum operators do not commute, and the choice
of convention affects the products of operators that arise upon transforming Eq. (\ref{Hrot}) to the molecule-fixed frame.
It is easy to show that the frame transformation adopted in this work leads to
$\hat{J}^\text{SF}\cdot\hat{S}^\text{SF} = \hat{J}^\text{MF}\cdot\hat{S}^\text{MF}$, whereas that of Ref.~\cite{GerritO2} leads to
$\hat{J}^\text{SF}\cdot\hat{S}^\text{SF} = \hat{S}^\text{MF} \cdot \hat{J}^\text{MF}$.
The matrix elements of $\hat{H}_\text{rot}$ are independent of the convention. In the following, we will omit the superscript ``MF''.

The second term in Eq. (\ref{Hmol}) is the spin-orbit (SO) interaction
\begin{equation}\label{HSO}
\hat{H}_\text{SO}=A\hat{L}\cdot\hat{S},
\end{equation} 
where $A$ is the SO interaction constant. The remaining terms in Eq. (\ref{Hmol}) account for the effects of $\Lambda$-doubling
and the interactions with static electric and magnetic fields (explicit expressions for $\hat{H}_\Lambda$, $\hat{H}_\text{E}$,
and $\hat{H}_\text{B}$ are given below).
The energy levels of a $^2\Pi$ molecule can be evaluated by diagonalizing the Hamiltonian (\ref{Hmol}) in Hund's case (a) basis
\begin{equation}\label{BasisU}
|JM\Omega\rangle |\Lambda\Sigma\rangle = \biggl{(}\frac{2J+1}{4\pi}\biggr{)}^{1/2}D^{J\star}_{M\Omega}(\bar{\alpha},\bar{\beta},0) |\Lambda\Sigma\rangle,
\end{equation} 
where $M$ and $\Omega$ are the projections of $\hat{J}$ onto the space-fixed and molecule-fixed quantization axes, 
$D^{J\star}_{M\Omega}$ is the Wigner $D$-function, and $|\Lambda\Sigma\rangle$ is the electronic wave function.
The molecule-fixed projections of $\hat{L}$ and $\hat{S}$ are denoted as $\Lambda$ and $\Sigma$. For a $^2\Pi$ electronic
state, they take the values $\Lambda = \pm1$ and $\Sigma = \pm \half$. The off-diagonal matrix elements of the SO interaction
between the $^2\Pi$ state and the excited electronic states give rise to the $\Lambda$-doubling effect \cite{Brown1,Brown2} described below.
After neglecting the cross terms, Eq. (\ref{HSO}) may be written as 
\begin{equation}\label{SOdiagonal}
\hat{H}_\text{SO} = A\hat{L}_z\hat{S}_z,
\end{equation} 
where the subscript $z$ refers to the molecule-fixed projections of the angular momentum operators.
The $\Lambda$-doubling is described by the effective Hamiltonian \cite{Brown1,Brown2}
\begin{equation}\label{LDgeneral}
\hat{H}_\Lambda = \half e^{-2i\phi} [-{q}\hat{J}_+^2 + ({p}+2{q})\hat{J}_+\hat{S}_+]
+\half e^{2i\phi} [-{q}\hat{J}_-^2 + ({p}+2{q})\hat{J}_-\hat{S}_-],
\end{equation} 
where $\hat{J}_\pm = \hat{J}_x \mp i\hat{J}_y$ and $\hat{S}_\pm = \hat{S}_x \pm i\hat{S}_y$ are the ladder operators,
$\phi$ is the azimuthal angle of the electron in the molecule-fixed frame, and ${p}$ and ${q}$ are the
phenomenological $\Lambda$-doubling parameters. Using the phase convention for the
electronic wave functions, $\langle \Lambda=\pm 1 | e^{\pm 2i\phi} | \Lambda' = \mp1\rangle = -1$,
the matrix elements of the Hamiltonian (\ref{LDgeneral}) can be written as
\begin{equation}\label{LD}
\langle \Lambda| \hat{H}_\Lambda |\Lambda'\rangle = \half \left( \delta_{\Lambda',-1}\delta_{\Lambda,1} [ {q}\hat{J}_+^2 -({p}+2{q})\hat{J}_+\hat{S}_+]
+ \delta_{\Lambda',1}\delta_{\Lambda,-1} [ {q}\hat{J}_-^2 -({p}+2{q})\hat{J}_-\hat{S}_- ] \right)
\end{equation} 
The interaction with the magnetic field of strength $B$ has the form
\begin{equation}\label{HB}
\hat{H}_\text{B} = \mu_0 B(\hat{L} + 2\hat{S})\cdot \hat{B},
\end{equation}
where $\hat{B}$ is the unit vector in the direction of the external magnetic field.
To first order, the interaction of the molecule with the dc electric field can be written as
\begin{equation}\label{HE}
\hat{H}_\text{E} = -Ed\cos\chi,
\end{equation}
where $\chi$ is the polar angle of the molecule in the space-fixed frame, $E$ is the electric field strength,
and $d$ is the permanent electric dipole moment of the molecule. Here, we assume that both the electric 
and magnetic fields are oriented along the space-fixed $z$-axis.
The more general case of crossed electric and magnetic fields is considered elsewhere \cite{Erik}.

It is convenient to use parity-adapted Hund's case (a) basis functions
\begin{equation}\label{BasisA}
|JM\bOmega \epsilon \rangle = \half \left\{ |JM\bOmega\rangle |\Lambda=1,\Sigma = \bOmega -1 \rangle
+ \epsilon(-)^{J-1/2} |JM{-\bOmega}\rangle |\Lambda=-1,\Sigma={-\bOmega}+1 \rangle\right\},
\end{equation}
where $\bOmega=|\Omega|$, and the parity index $\epsilon=\pm1$ characterizes the inversion symmetry of the basis functions [$\epsilon(-)^{J-1/2}=1$
for $e$-parity states and $-1$ for $f$-parity states]. Note that in the parity-adapted basis, $\bOmega>0$ and the quantum
number $\Lambda$ does not have a definite value.
Expanding the rotational kinetic energy in terms of the ladder operators and using Eq. (\ref{SOdiagonal}), we obtain
\begin{equation}\label{HrotExp2}
\hat{H}_\text{rot} +\hat{H}_\text{SO}= B_e [\hat{J}^2 - 2\hat{J}_z^2 - \hat{J}_+\hat{S}_- - \hat{J}_-\hat{S}_+ + (A+2B_e)\hat{L}_z \hat{S}_z],
\end{equation} 
where we have omitted the terms $\hat{L}^2$ and $\hat{S}^2$ which would only result in an overall energy shift.
The matrix elements of the rotational and spin-orbit Hamiltonians can now be evaluated in the parity-adapted basis (\ref{BasisA}).
They have the form
\begin{multline}\label{HrotME} 
\langle JM\bOmega\epsilon | \hat{H}_\text{rot} +\hat{H}_\text{SO} | J'M'\bOmega'\epsilon'\rangle =\delta_{\epsilon\epsilon'}\delta_{JJ'}\delta_{MM'}
\{  B_e [J(J+1) - 2\bOmega^2 ]\delta_{\bOmega \bOmega'} + (A+2B_e)(\bOmega-1)\delta_{\bOmega \bOmega'} \\ 
- B_e [\delta_{\bOmega,\bOmega'-1} \alpha_{-} (J',\bOmega') \alpha_{-}(S,\bOmega'-1) 
- \delta_{\bOmega,\bOmega'+1} \alpha_{+}(J',\bOmega') \alpha_+(S,\bOmega'-1)] \},
\end{multline} 
where $\alpha_\pm(J,\bOmega) = \sqrt{J(J+1) - \bOmega(\bOmega\pm1)}$. Combining Eqs. (\ref{LD}) and (\ref{BasisA}),
we obtain the following compact expression for the $\Lambda$-doubling matrix elements 
\begin{align}\label{LDME} \notag
\langle JM\bOmega\epsilon | \hat{H}_\Lambda | J'M'\bOmega'\epsilon'\rangle = \half\delta_{\epsilon\epsilon'}\delta_{JJ'}\delta_{MM'}
\epsilon (-)^{J-1/2} & [{q}\alpha_-(J',\bOmega')\alpha_-(J',\bOmega'-1) \delta_{\bOmega',2-\bOmega'} \\
&- ({p}+2{q})\alpha_-(J',\bOmega')\alpha_+(S,\bOmega'-1) \delta_{\bOmega,1-\bOmega'}].
\end{align} 
In order to evaluate the matrix elements of the Zeeman Hamiltonian in the basis (\ref{BasisA}), it is necessary to transform the
operator (\ref{HB}) to the molecule-fixed frame
\begin{equation}\label{HrotExp}
\hat{H}_\text{B} = \mu_0 B \sum_{q=-1}^1 (\hat{L}_q + 2\hat{S}_q) D^{1\star}_{0q} (\bar{\alpha},\bar{\beta},0),
\end{equation} 
where only the $q=0$ molecule-fixed component of $\hat{L}$ survives on the right-hand side.
Evaluating the integrals over the product of three Wigner $D$-functions, we find
\begin{align}\label{HBME} \notag
\langle JM\bOmega\epsilon | \hat{H}_\text{B} | J'M'\bOmega'\epsilon'\rangle = \mu_0 B \delta_{\epsilon\epsilon'}\delta_{MM'}(-)^{M'-\Omega'}
[(2J+1)(2J'+1)]^{1/2}\threejm{J}{M}{1}{0}{J'}{-M'} \\ \notag \times 
\biggl{[} \sqrt{2}\alpha_+(S,\bOmega'-1)\threejm{J}{\bOmega}{1}{-1}{J'}{-\bOmega'}
- \sqrt{2}\alpha_-(S,\bOmega'-1)\threejm{J}{\bOmega}{1}{1}{J'}{-\bOmega'} \\
+ (2\bOmega-1)\threejm{J}{\bOmega}{1}{0}{J'}{-\bOmega'}.
\biggr{]}
\end{align} 
The matrix elements of the interaction with electric fields have a similar form
\begin{align}\label{HEME} \notag
\langle JM\bOmega\epsilon | \hat{H}_\text{E} | J'M'\bOmega'\epsilon'\rangle = -Ed \delta_{\epsilon,-\epsilon'}\delta_{MM'}(-)^{M'-\Omega'}
[(2J+1)(2J'+1)]^{1/2}  \\  \times \threejm{J}{M}{1}{0}{J'}{-M'}  \threejm{J}{\bOmega}{1}{0}{J'}{-\bOmega'}.
\end{align} 
This expression shows that electric fields couple the states of the opposite inversion parity.

\subsection{Collision dynamics}

The He--OH($^2\Pi$) collision complex can be described by the Jacobi vectors $\bm{R}$ - the separation of He
from the center of mass of OH and $\bm{r}$ - the internuclear distance in OH. The angle between the vectors is denoted by $\theta$.
In the following, it will be convenient to use the unit vectors $\hat{R}=\bm{R}/R$ and $\hat{r}=\bm{r}/r$,
where $R=|\bm{R}|$, $r=|\bm{r}|$. The Hamiltonian of the collision complex can be written in atomic units as \cite{Gianturco,MillardJCP,Bohn3}
\begin{equation}\label{H}
\hat{H} = -\frac{1}{2\mu R}\frac{\partial^2}{\partial R^2}R + \frac{\hat{\ell}^2}{2\mu R^2} + \hat{V}(R,r,\theta) + \hat{H}_\text{mol},
\end{equation} 
where $\mu$ is the reduced mass of the $^3$He--OH system, $\hat{\ell}$ is the orbital angular momentum for the collision,
$\hat{V}(R,r,\theta)$ is the electrostatic interaction potential, and  $\hat{H}_\text{mol}$ is the Hamiltonian of the OH molecule in the presence
of external electric and magnetic fields (see Sec. IIA).
We assume that the internuclear distance of OH is fixed at the equilibrium value of 1.226~\AA.
The wave function of the He--OH collision complex $\Psi$ satisfies the Schr{\"o}dinger equation at a total energy $E$,
and can be expanded over the complete coupled-channel basis 
\begin{equation}\label{PsiExp}
\Psi = \frac{1}{R} \sum_\beta F_\beta(R) \psi_\beta (\hat{R},\hat{r}),
\end{equation} 
where $\psi_\beta (\hat{R},\hat{r})$ are the angular basis functions.
The fully uncoupled angular basis set can be defined as a direct product of the parity-unadapted Hund's case (a)
functions and spherical harmonics
\begin{equation}\label{basisI} 
|JM\Omega\rangle |\Lambda\Sigma\rangle |\ell m_\ell\rangle,
\end{equation} 
where the spherical harmonics $|\ell m_\ell\rangle = Y_{\ell m_\ell}(\hat{r})$ describe the orbital motion of the
He atom around the OH fragment. To be consistent with spectroscopic nomenclature, it is convenient
to use a slightly modified basis given by
\begin{equation}\label{basisII} 
|JM\bOmega\epsilon\rangle |\ell m_\ell\rangle,
\end{equation} 
where $|JM\bOmega\epsilon\rangle$ are Hund's case (a) basis functions of definite parity (\ref{BasisA}).
The basis sets (\ref{basisI}) and (\ref{basisII}) are related by a unitary transformation, and are equivalent.

Substituting the coupled-channel expansion (\ref{PsiExp}) into the Schr{\"o}dinger equation,
we obtain a system of coupled second-order differential equations
\begin{equation}\label{CCsystem} 
\left[ \frac{d^2}{dR^2} + 2\mu E\right] F_{\beta}(R) = 2\mu\sum_{\beta'}\langle \psi_{\beta}(\hat{R},\hat{r}) | \hat{V}(R,\theta)
+ \frac{\hat{\ell}^2}{2\mu R^2} +\hat{H}_\text{mol} |\psi_{\beta'}(\hat{R},\hat{r})\rangle F_{\beta'}(R).
\end{equation} 
In order to solve these equations, it is necessary to evaluate the matrix elements on the right-hand side.
In the uncoupled representation (\ref{basisII}), the operator $\ell^2$ is diagonal with matrix elements given by $\ell(\ell+1)$ \cite{Roman2004}.
The matrix elements of the asymptotic Hamiltonian $\hat{H}_\text{mol}$ in basis (\ref{basisII}) are
\begin{equation}\label{CCsystem2} 
\langle JM\bOmega \epsilon | \langle \ell m_\ell| \hat{H}_\text{mol}| J'M'\bOmega' \epsilon'\rangle |\ell' m_\ell'\rangle
=\delta_{\ell \ell'}\delta_{m_\ell m_\ell'}\langle JM\bOmega \epsilon |\hat{H}_\text{mol}| J'M'\bOmega' \epsilon'\rangle,
\end{equation} 
where the expression on the right-hand side is evaluated in Sec. IIA. All that remains to complete the definition of the system
of coupled equations (\ref{CCsystem}) is to evaluate the matrix elements of the interaction potential. This is described in the following section.

Once the coupled equations are solved, the asymptotic wave function
is transformed to the field-dressed basis $|\gamma\rangle |\ell m_\ell\rangle$, which diagonalizes the asymptotic
Hamiltonian $\hat{H}_\text{mol}$. The transformation can be written as
\begin{equation}\label{dressed} 
|\gamma\rangle |\ell m_\ell\rangle = |\ell m_\ell\rangle \sum_{JM\Omega\epsilon}C_{JM\bOmega\epsilon,\gamma}|JM\bOmega\epsilon\rangle,
\end{equation} 
where $C_{JM\bOmega\epsilon,\gamma}$ are the components of the eigenvector of $\hat{H}_\text{mol}$ corresponding
to the eigenstate $\gamma$ with energy $\varepsilon_\gamma$. The matrix of the transformation (\ref{dressed}) is diagonal in $\ell$ and $m_\ell$.
The $S$-matrix can be obtained from the transformed wave function using the standard asymptotic matching procedure \cite{Johnson}.
The cross sections for transitions between the field-dressed states of OH can be expressed as
\begin{equation}\label{xs} 
\sigma_{\gamma\to \gamma'} = \frac{\pi}{k_\gamma^2}\sum_{M_\text{tot}} \sum_{\ell,m_\ell} \sum_{\ell',m_\ell'}|\delta_{\gamma\gamma'}
\delta_{\ell\ell'}\delta_{m_\ell,m_\ell'} - S^{M_\text{tot}}_{\gamma \ell m_\ell;\gamma'\ell'm_\ell'}|^2,
\end{equation} 
where the wavevector $k^2_\gamma = 2\mu(E-\varepsilon_\gamma)=2\mu E_\text{coll}$, $E_\text{coll}$ is the collision energy,
and the summation in Eq. (\ref{xs}) is performed in a cycle over the total angular momentum projection~\cite{Roman2004}.

We used the following molecular constants of OH (in cm$^{-1}$): $B_e=18.55$,
$A = -139.273$, ${p} = 0.235608$, ${q}=-0.03877$ \cite{Brown1,Bohn3}. The permanent electric dipole moment
of OH($^2\Pi$) $d =1.68$ D was taken from Ref. \cite{Bohn3}. 
The coupled-channel expansion (\ref{PsiExp}) included all basis states with $J\leq \frac{11}{2}$ and $\ell\leq 5$, which resulted in a total
of 622 coupled channels for $M_\text{tot}=\frac{1}{2}$. The close-coupled equations (\ref{CCsystem}) were solved
numerically using the improved log-derivative method \cite{David} on a grid of $R$ from 2 to 65 $a_0$ with a step size of 0.01 $a_0$.
The resulting cross sections were converged to within 5\%.

\subsection{Matrix elements of the interaction potential}

The matrix elements of the interaction potential between the states with definite $\Lambda$ can be expanded in
reduced Wigner $D$-functions 
\begin{align}\label{PESexp} 
V_{\Lambda\Lambda'}(R,\theta) = \sum_{\lambda} D^{\lambda\star}_{0,\Lambda'-\Lambda}(0,\theta, 0)V_{\lambda,\Lambda'-\Lambda}(R).
\end{align} 
Since in our case $|\Lambda|=|\Lambda'|=1$, only the terms in Eq. (\ref{PESexp}) with $\Lambda'-\Lambda=0,\pm 2$ are different from zero.
They can be obtained by expanding the half-sum and half-difference
of the two ground-state potential energy surfaces of $A'$ and $A''$ symmetry \cite{MillardJCP,MillardCP}
\begin{align}\label{PESexp2} 
\half(V_{A'}+V_{A''}) &= \sum_{\lambda=0}^{\lambda_\text{max}} P_\lambda (\cos\theta)V_{\lambda 0}(R)  \\
\half(V_{A''}-V_{A'}) &= \sum_{\lambda = 2}^{\lambda_\text{max}} d^\lambda_{02}(\cos\theta) V_{\lambda 2}(R)
\end{align} 
where $d^\lambda_{0\mu}(\cos\theta)$ are the reduced Wigner $D$-functions and $P_{\lambda \mu}(\cos\theta)$
are the associated Legendre polynomials, which are related through \cite {Zare}
\begin{equation}\label{relation} 
d^\lambda_{0\mu}(\cos\theta) = \left[ \frac{(\lambda-\mu)!}{(\lambda+\mu)!} \right]^{1/2} P_{\lambda \mu}(\cos\theta).
\end{equation} 
The interaction potential (\ref{PESexp}) can be evaluated in the parity-unadapted basis using Eqs. (\ref{BasisU}), (\ref{basisI})
and the generalized spherical harmonics addition theorem \cite{MillardCP,Zare}
\begin{equation}\label{SHAT} 
d^\lambda_{0\mu}(\cos\theta) = (-)^{-\mu}d^\lambda_{0,-\mu}(\cos\theta)
=(-)^{\mu}\sum_{m_\lambda} \left[ \frac{4\pi}{2\lambda+1}\right]^{1/2}
D^{\lambda}_{m_\lambda\mu}(\bar{\alpha},\bar{\beta},0) Y_{\lambda m_\lambda}(\hat{R}).
\end{equation} 
Combining this expression with Eq. (\ref{PESexp}) and evaluating the integrals over the products of three $D$-functions,
we obtain
\begin{multline}\label{MEbasisI} 
\langle JM\Omega |\langle \Lambda \Sigma |\langle \ell m_\ell | \hat{V}(R,\theta) | J'M'\Omega'\rangle | \Lambda'\Sigma'\rangle |\ell'm_\ell'\rangle
= \delta_{\Sigma\Sigma'} [(2J+1)(2J'+1)(2\ell+1)(2\ell'+1)]^{1/2} \\ \times (-)^{m_\ell +M'-\Omega'}
\sum_{\lambda,m_\lambda}\threejm{J}{M}{\lambda}{m_\lambda}{J'}{-M'} \threejm{J}{\Omega}{\lambda}{\Lambda'-\Lambda}{J'}{-\Omega'}
\threejm{\ell}{-m_\ell}{\lambda}{m_\lambda}{\ell'}{m_\ell'}
\threejm{\ell}{0}{\lambda}{0}{\ell'}{0} \\ \times V_{\lambda,\Lambda'-\Lambda}(R).
\end{multline}
The expansion coefficients have the property $V_{\lambda,\Lambda'-\Lambda}(R)=V_{\lambda,\Lambda-\Lambda'}(R)$ \cite{MillardCP}.
We note that the 3-$j$ symbols in Eq. (\ref{MEbasisI}) vanish unless $m_\lambda = M'-M = m_\ell-m_\ell'$. Thus, the electrostatic
interaction potential only couples the states with the same total angular momentum projection $M_\text{tot}=M+m_\ell = M' +m_\ell'$.
A transformation of the interaction potential matrix elements (\ref{MEbasisI}) to the parity-adapted basis (\ref{BasisA}) using
the symmetry properties of 3-$j$ symbols \cite{Zare} yields
\begin{multline}\label{MEbasisII} 
\langle JM\bOmega\epsilon| \langle \ell m_\ell | \hat{V}(R,\theta) | J'M'\bOmega'\epsilon'\rangle |\ell'm_\ell'\rangle
= [(2J+1)(2J'+1)(2\ell+1)(2\ell'+1)]^{1/2} (-)^{m_\ell +M'-\bOmega'}  \\ \times
\sum_{\lambda,m_\lambda} \frac{1}{2}[1 + \epsilon\epsilon'(-)^\lambda]  \threejm{J}{M}{\lambda}{m_\lambda}{J'}{-M'}
\threejm{\ell}{-m_\ell}{\lambda}{m_\lambda}{\ell'}{m_\ell'}
\threejm{\ell}{0}{\lambda}{0}{\ell'}{0}  \\ \times
\left[ \delta_{\bOmega\bOmega'} \threejm{J}{\bOmega}{\lambda}{0}{J'}{-\bOmega'}V_{\lambda 0}(R) 
-\epsilon'(-)^{J'-1/2} \delta_{\bOmega,2-\bOmega'} \threejm{J}{\bOmega}{\lambda}{-2}{J'}{\bOmega'}V_{\lambda 2}(R)  \right]
\end{multline}
An analysis of the expression in square brackets shows that the rotational levels in the same SO manifold
($\bOmega'=\bOmega$) are coupled by the half-sum PES (\ref{PESexp2}), whereas the levels belonging to different SO manifolds
($\bOmega'=2-\bOmega$) are coupled by the half-difference PES.
Similarly, the factor $\half[1+\epsilon\epsilon'(-)^\lambda]$ ensures that the couplings between the states of the same
parity ($\epsilon=\pm 1\leftrightarrow \epsilon'=\pm 1$) are induced by the anisotropic terms with even $\lambda$, and those of
the states of opposite parity ($\epsilon=\pm 1\leftrightarrow \epsilon'=\mp 1$) are induced by the anisotropic terms with odd~$\lambda$.

\section{Results}

\begin{figure}[t]
	\centering
	\includegraphics[width=0.5\textwidth, trim =0 20 0 90]{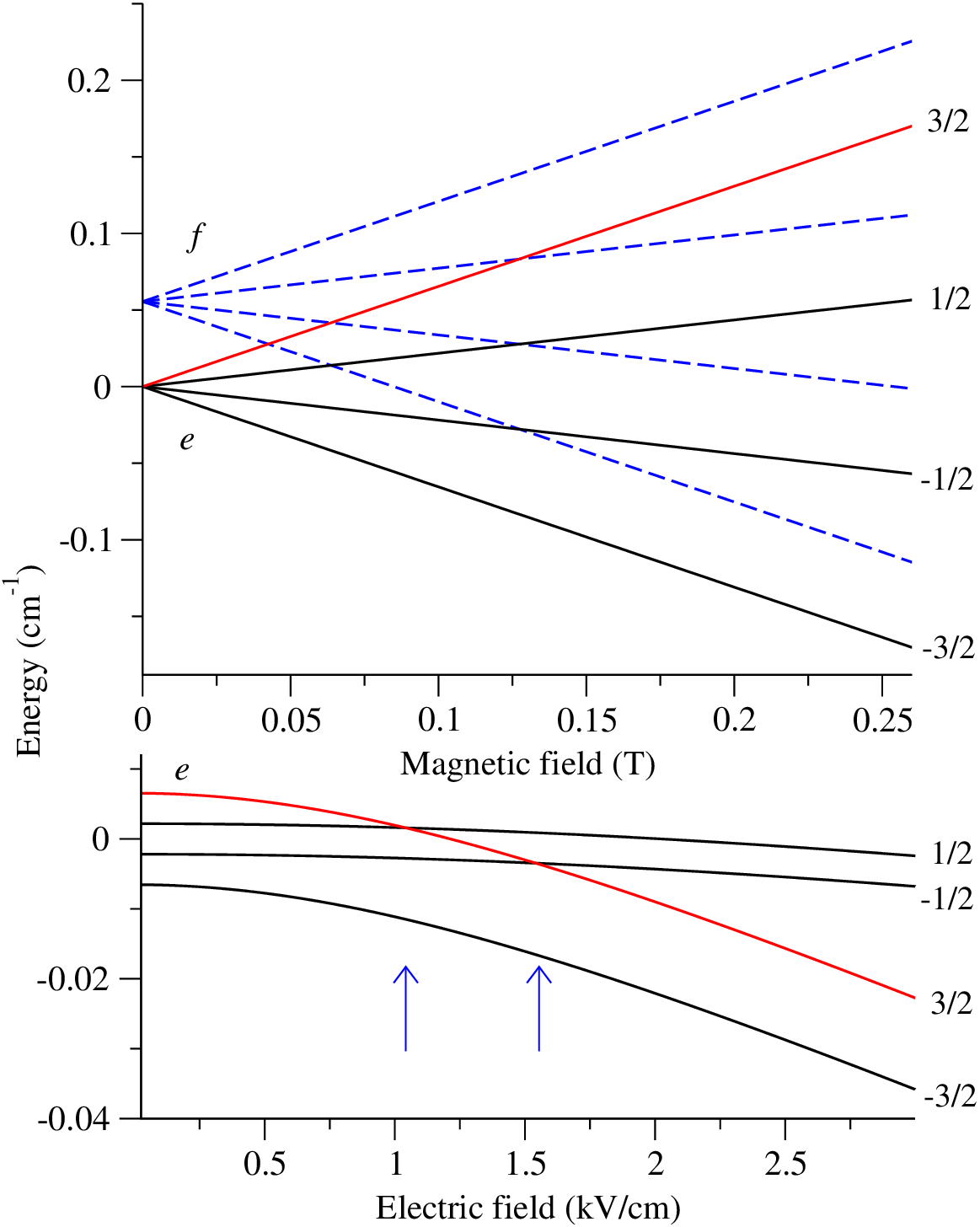}
	\renewcommand{\figurename}{Figure}
	\caption{Zeeman energy levels of OH. The initial state for scattering calculations (see text) is denoted by the red (light grey) solid line.
	The values of $M$ for individual magnetic sublevels are shown to the right of the curves.}
\end{figure}

The upper panel of Fig. 1 shows the energy levels of OH as functions of the magnetic field. At zero field, 
the absolute ground state of the molecule is a $\Lambda$-doublet with $J=3/2$. Magnetic fields further split
the $e$ and $f$ components of the doublet into four Zeeman levels characterized by $M=-\frac{3}{2},-\frac{1}{2},\frac{1}{2},\frac{3}{2}$
(in order of increasing energy). As mentioned above, $M$ is rigorously conserved in parallel fields and we will use this label to
classify the molecular states.
The magnetic levels corresponding to different manifolds cross at $B\sim 0.1$ T, where the Zeeman shift becomes
comparable to the $\Lambda$-doublet splitting. The matrix elements of the Zeeman Hamiltonian (\ref{HBME}) are diagonal in $\epsilon$
and independent of its sign. Therefore, the $e$ and $f$ Zeeman manifolds are identical and the crossings between the levels
of different manifolds are not avoided, as illustrated in the upper panel of Fig. 1. Electric fields couple the opposite parity states,
leading to avoided crossings similar to those encountered in $^2\Sigma$ molecules
\cite{OurWork1,Erik}. An important difference is
that crossings in $^2\Pi$ molecules occur at small magnetic fields $B\sim 0.1$ T corresponding to typical $\Lambda$-doublet
splittings of tenths of cm$^{-1}$. In contrast, the electric-field induced crossings in $^2\Sigma$ molecules occur between different rotational
levels, which requires magnetic fields on the order of several Tesla \cite{OurWork1,OurWork2}.

The lower panel of Fig. 1 shows the electric field dependence of the $e$-manifold energy levels in the presence
of a static magnetic field of 0.01 T. The four $e$-states shift downwards with increasing the field, whereas the corresponding
$f$-states (not shown) shift in the opposite direction. It follows from Eq. (\ref{HEME}) that the $|M|=\frac{3}{2}$ states
have larger $g$-factors, so their energy decreases faster than that of the $|M|=\frac{1}{2}$ states. As a result, two avoided
crossings occur at electric fields of about 1 and 1.5 kV/cm shown by the vertical arrows in the lower panel of Fig. 1. Note that
the location of the crossings depends on magnetic field strength, shifting to higher electric fields with increasing magnetic field.
In the following, we will consider collisions of OH molecules initially in the state $|J=\frac{3}{2},M=\frac{3}{2},\epsilon=-1\rangle$
denoted by the red (light grey) line in Fig. 1. The OH molecules selected in this state gain potential energy with increasing magnetic
field and can be confined in a permanent magnetic trap. The expansion of the initial state in terms of Hund's case (a) basis functions
(\ref{dressed}) is
\begin{align}\label{initial}\notag
|J=\textstyle\frac{3}{2},M=\frac{3}{2},\epsilon=-1\rangle &= 0.985 |J=\textstyle\frac{3}{2},\bOmega=\frac{3}{2},M=\frac{3}{2},\epsilon=-1\rangle
\\ &+ 0.174 |J=\textstyle\frac{3}{2},\bOmega=\frac{1}{2},M=\frac{3}{2},\epsilon=-1\rangle.
\end{align}
This equation illustrates that OH is not a pure Hund's case (a) molecule: different $\bOmega$ components of
the basis (\ref{BasisA}) are mixed by the cross terms $\hat{J}_+\hat{S}_-$ and $\hat{J}_- \hat{S}_+$ in Eq. (\ref{HrotExp2}).
Because the rotational constant of OH is large compared to Zeeman splittings,
the mixing coefficients in Eq. (\ref{initial}) are independent of $M$ and magnetic field.
In the presence of an electric field, the initial state (\ref{initial}) contains an admixture of basis functions of opposite
parity ($\epsilon=1$), whose contribution increases linearly with the field strength.
The energy of the state given by Eq. (\ref{initial}) decreases with increasing electric field 
as shown in the lower panel of Fig. 1.

\begin{figure}[t]
	\centering
	\includegraphics[width=0.6\textwidth, trim =0 20 0 40]{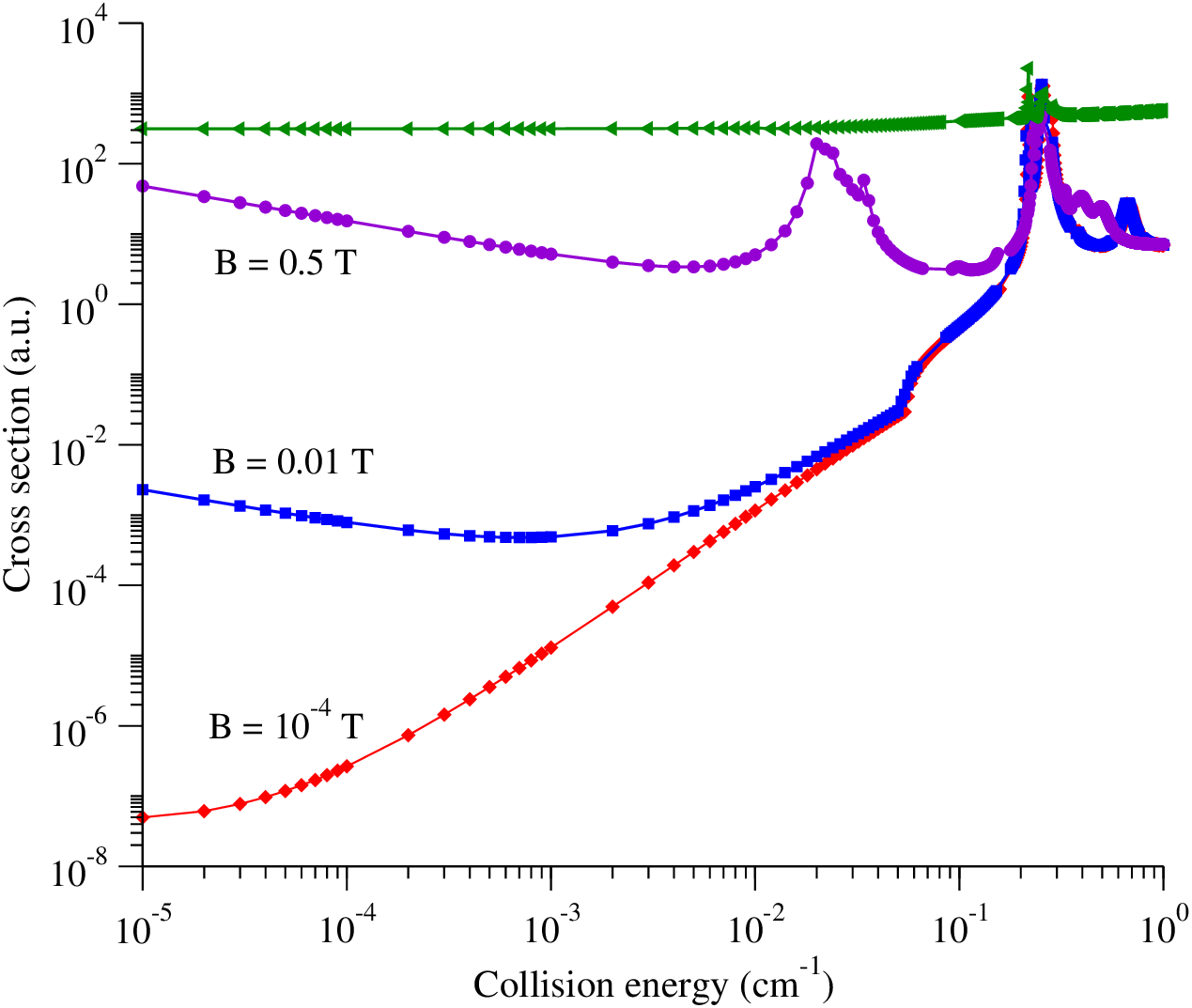}
	\renewcommand{\figurename}{Figure}
	\caption{Cross sections for elastic scattering and spin relaxation in $^3$He--OH collisions as functions of collision energy
	at different magnetic fields indicated in the graph. The elastic cross section is shown for $B=10^{-4}$ T.}
\end{figure}

Figure 2 shows the cross sections for elastic scattering and inelastic spin relaxation in $^3$He--OH collisions as functions of collision energy.
The behavior of the cross sections in the $E_\text{coll}\to 0$ limit is dictated by the Wigner threshold laws \cite{Wigner}:
the inelastic cross section increases as $E_\text{coll}^{-1/2}$ and the elastic cross section is independent of $E_\text{coll}$.
Magnetic fields modify the energy dependence of the cross sections. At small magnetic fields, the 
cross sections continue to decrease with $E_\text{coll}$ down to $10^{-5}$ K and start to follow the
threshold behavior as the energy is further decreased. The turnover point moves to higher energy with increasing
magnetic field. At very large fields, the spin relaxation cross section always increases with decreasing collision energy.
This behavior in qualitatively similar to that observed for collisions of molecules in $\Sigma$ electronic states \cite{Roman2004}.
The conservation of the total angular momentum projection (see Sec. IIB) implies that the spin relaxation transition
$|M=\frac{3}{2}\rangle\to |M'=\frac{1}{2}\rangle$ should be accompanied by the transition $m_\ell\to m_\ell'=m_\ell+1$
which leads to a centrifugal barrier in the outgoing collision channel.
The centrifugal barrier suppresses inelastic processes as long as the energy defect between the initial and final
Zeeman levels does not exceed the barrier height \cite{Roman2004,VolpiBohn}. At higher magnetic fields (or collision energies),
the centrifugal barrier is easily surmounted and spin relaxation rates increase dramatically, as illustrated in Fig. 2.

An interesting feature apparent in Fig. 2 is the rapid increase of spin relaxation as the collision energy is varied through the
$\Lambda$-doubling threshold (0.06 cm$^{-1}$). This is caused by the opening of new relaxation channels in the
higher-energy $f$-manifold (see the upper panel of Fig. 1). At collision energies above
0.1 K, both elastic and inelastic cross sections display a rich resonance structure. The resonance pattern is rather dense
and features both shape and Feshbach resonances. This is in contrast with $\Sigma$-state molecules where a single
(or at most several) shape resonances are typically present \cite{OurWork1,Erik}. The resonances grow in number with increasing
magnetic field. At $B=0.5$ T, the spin relaxation cross section shows two distinct peaks. We attribute the peaks to shape
resonances in the outgoing collision channels which are split by the magnetic field. A similar separation of shape resonances
has been observed for NH($^3\Sigma^-$)--He collisions \cite{NHisotopes}.
From Fig. 2, the ratio of the cross sections for elastic scattering and spin relaxation varies from 1 to 100
in the temperature interval 0.01 - 1 K. Therefore, cryogenic cooling and magnetic trapping of OH using $^3$He buffer gas would be
extremely challenging. The rate constant for spin relaxation is $1.2\times 10^{-11}$ cm$^3$/s at $T=0.1$ K and $B=0.01$ T,
which corresponds to the OH trapping lifetime of $\sim$0.1 ms at the buffer gas density of 10$^{15}$ cm$^{-3}$.
Although spin relaxation is suppressed at collision energies below 10 mK and magnetic fields $< 0.01$ T, this regime
is far beyond capability of modern cryogenic cooling techniques.

\begin{figure}[t]
	\centering
	\includegraphics[width=0.6\textwidth, trim =0 20 0 20]{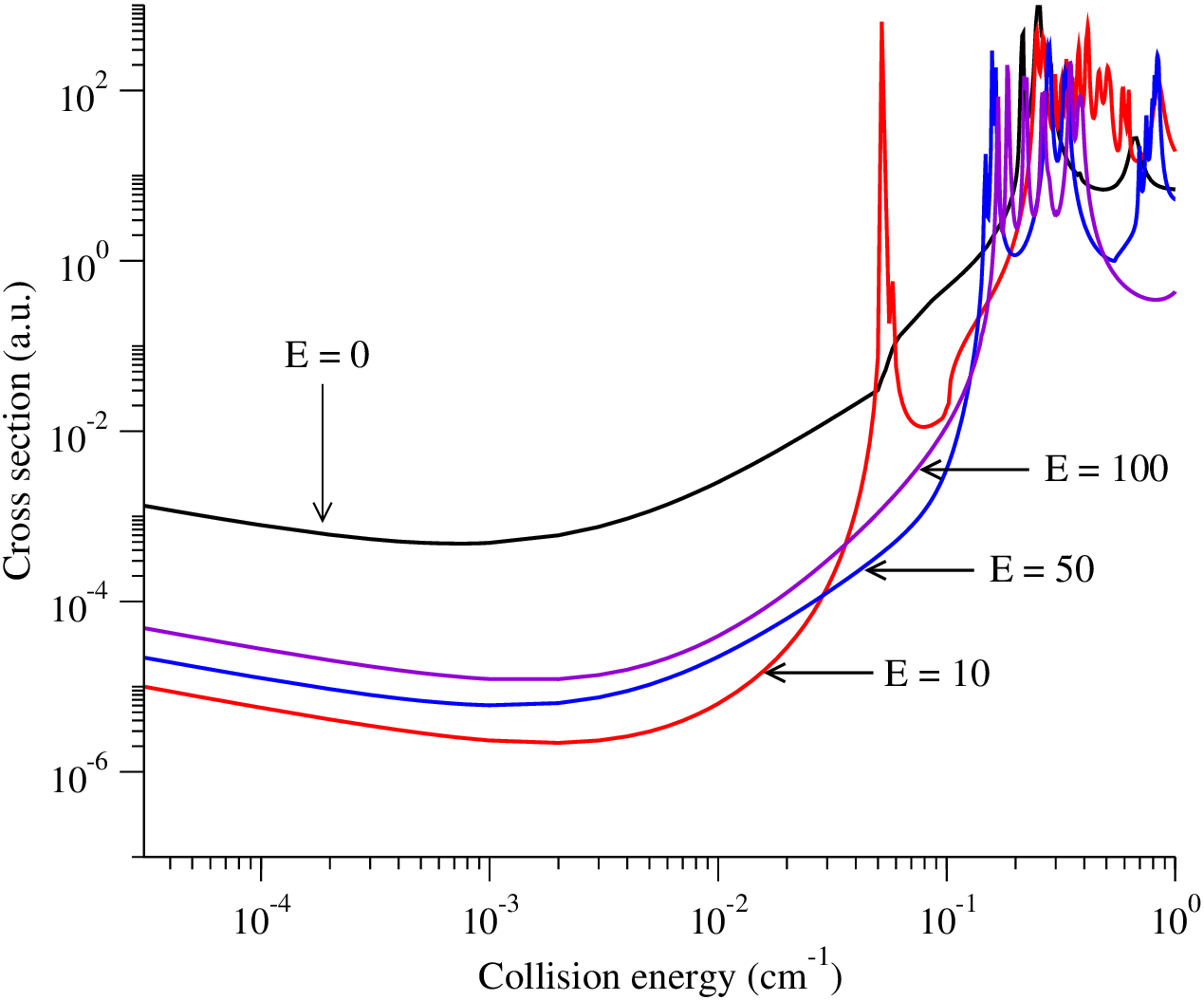}
	\renewcommand{\figurename}{Figure}
	\caption{Cross sections for elastic scattering and spin relaxation in $^3$He--OH collisions as functions of collision energy
	calculated for several electric field strengths (in kV/cm) indicated in the graph. The magnetic field is fixed at 0.01 T.}
\end{figure}

Equations (\ref{MEbasisII}) and (\ref{initial}) establish that different Zeeman levels of OH are directly coupled by the
atom-molecule interaction potential. Therefore, spin relaxation in collisions of $^2\Pi$ molecules with $^1S_0$
atoms is a direct process, in which all Zeeman states get mixed up in a collision mediated by electrostatic interactions.
The interaction potential for molecules in $\Sigma$ electronic states is diagonal in spin degrees
of freedom, and spin-changing transitions in $^2\Sigma$ molecules occur via a two-step mechanism through
the coupling of the ground and the first excited rotational states and the spin-rotation interaction \cite{Roman2004}.
Collision-induced spin relaxation in $^3\Sigma$ molecules follows a similar mechanism involving the spin-spin interaction. In the case of OH,
direct couplings of different magnetic sublevels arise due to the anisotropic terms ($\lambda >0$) in the expansion of the
interaction potential over the Wigner $D$-functions (\ref{PESexp}).

Figure 3 shows the cross sections for spin relaxation as a function of collision energy at selected values of the electric field.
Electric fields suppress spin relaxation in the ultracold regime. The origin of this effect is explained below.
At collision energies larger than 0.1 cm$^{-1}$, the suppression is much less efficient.
Another interesting effect is shifting and splitting of scattering resonances by electric fields.
The suppression of shape resonances is caused by the electric field-induced mixing of different partial waves \cite{JCP}.
The number of Feshbach resonances grows as the electric field is increased from zero
to 10 kV/cm, and the resonances shift to lower energies. We attribute this to the electric field-induced couplings between
the opposite parity states, which are uncoupled in the absence of the field, leading to additional avoided
crossings and Feshbach resonances.
The results shown in Fig. 3 suggest that collisional spin relaxation of OH at temperatures below 0.01 K can be suppressed
by two orders of magnitude with electric fields on the order of 10 kV/cm. The suppression is most efficient at low collision
energies. The He--OH spin relaxation rate in the absence of an electric field is $4.8\times 10^{-16}$ cm$^3$/s at $T=0.01$ K and $B=0.01$ T.
This value decreases to $4.4\times 10^{-18}$ in an electric field of 50 kV/cm. At 0.1 K, the rates are $1.1\times 10^{-11}$ and
$3.4\times 10^{-12}$ cm$^3$/s, respectively. Therefore, cryogenic cooling of magnetically
trapped OH may be greatly facilitated in the presence of an electric field. A new electromagnetic trap 
for OH molecules \cite{OHtrap} should be particularly suitable for experimental demonstration of the electric field-enhanced
evaporative cooling.

\begin{figure}[t]
	\centering
	\includegraphics[width=0.6\textwidth, trim =0 20 0 20]{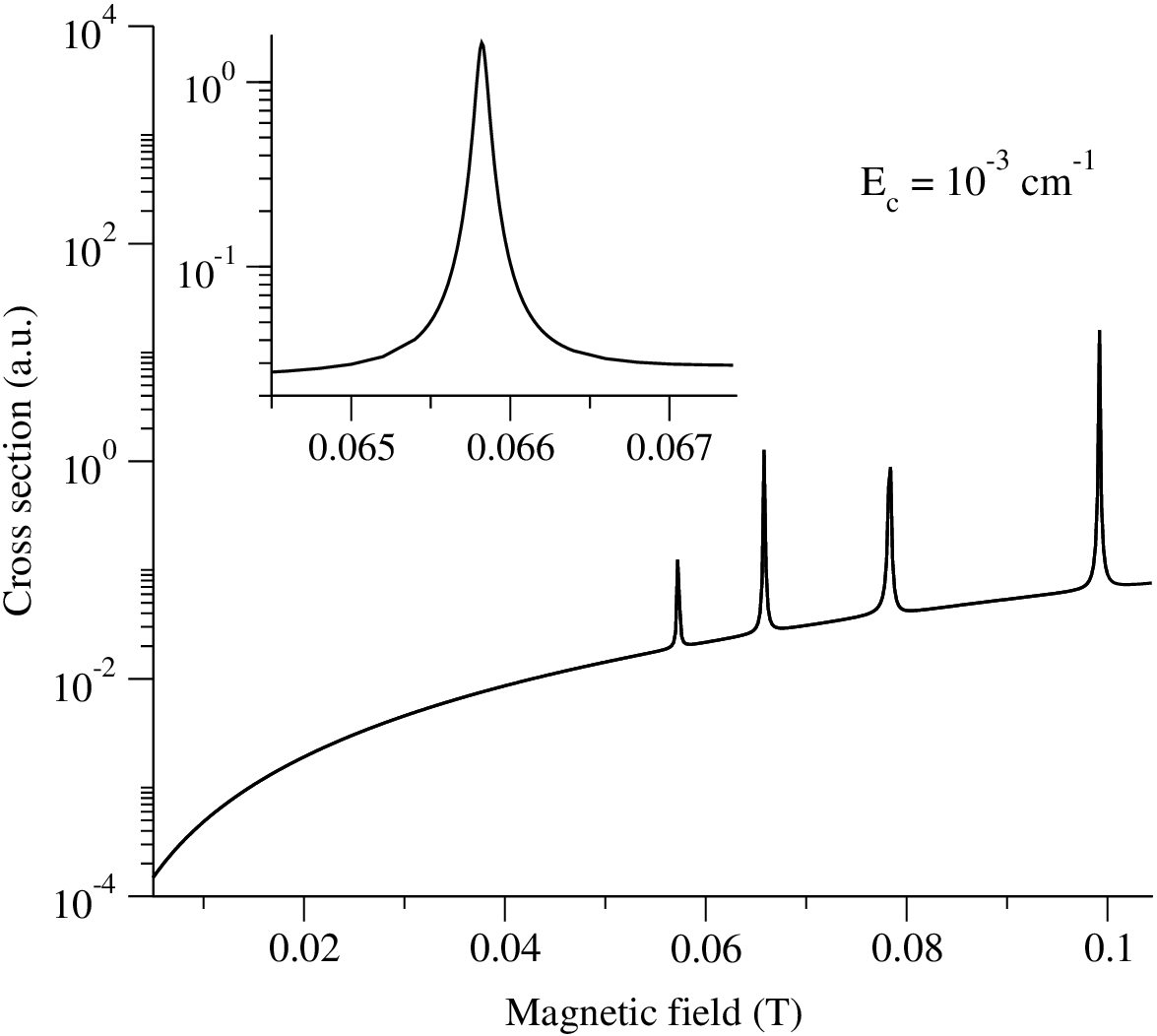}
	\renewcommand{\figurename}{Figure}
	\caption{Magnetic field dependence of spin relaxation cross sections at $E_\text{coll}=10^{-3}$ cm$^{-1}$ in
	the absence of an electric field.}
\end{figure}

In Fig. 4, we show the magnetic field dependence of spin relaxation cross sections at a collision energy of
10$^{-3}$ cm$^{-1}$ and zero electric field. The cross section shows sharp resonances superimposed on
a smoothly varying background. The Feshbach resonances arise due to the coupling of the initial channel
$|J=\frac{3}{2},M=\frac{3}{2},\epsilon=-1\rangle$ (\ref{initial}) with the closed channels $|J'=\frac{3}{2},M',\epsilon=1\rangle$
induced by the interaction potential. The closed channels are the bound states of the He$\cdots$OH van der Waals
complex that correlate to the upper Zeeman manifold in Fig. 1 in the limit $R\to\infty$. The inset in Fig. 4 demonstrates
that resonances may lead to a 100-fold enhancement of spin relaxation cross sections at certain magnetic fields.
The elastic cross section (not shown in Fig. 4) is not affected by Feshbach resonances.

\begin{figure}[t]
	\centering
	\includegraphics[width=0.6\textwidth, trim =0 20 0 20]{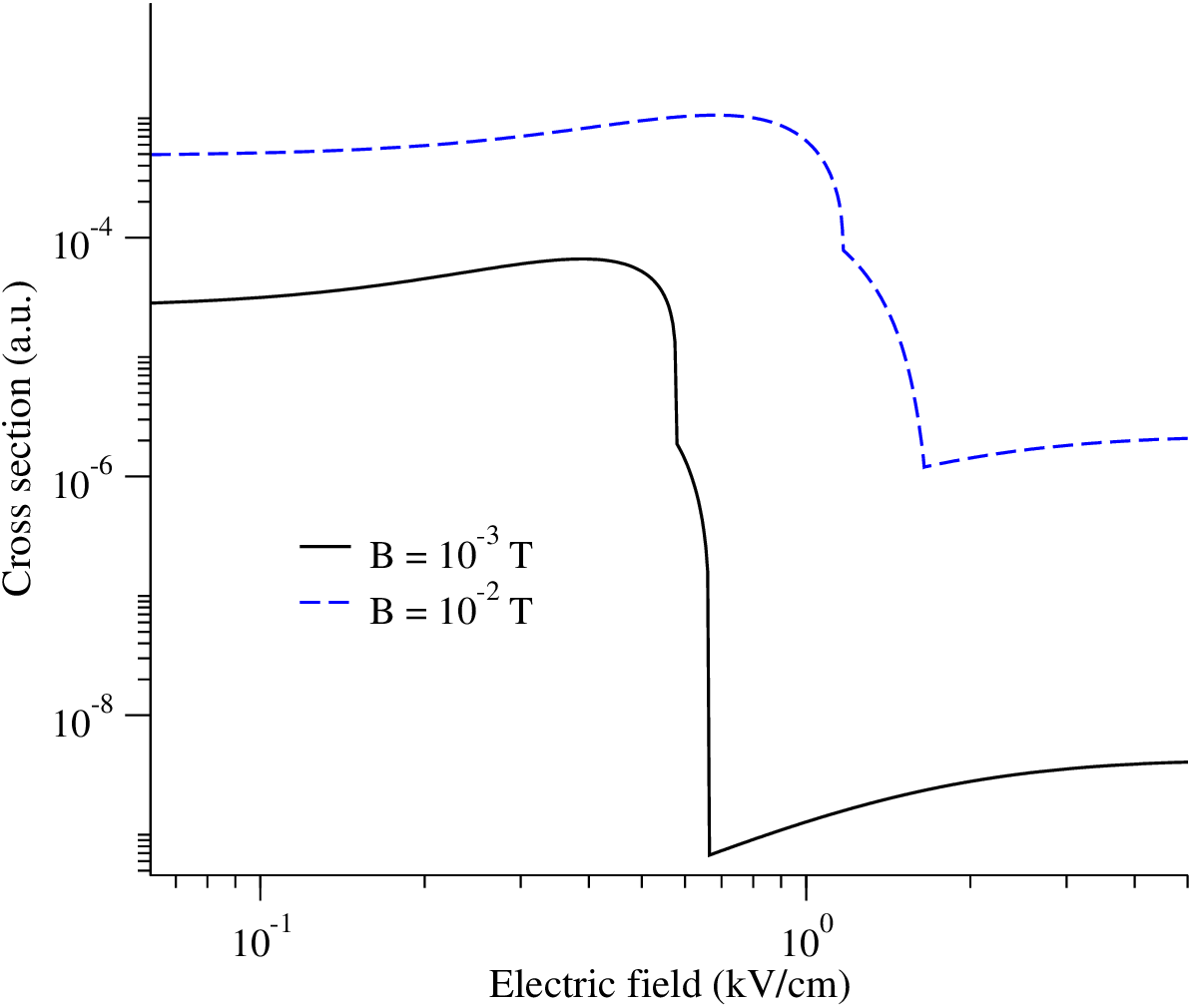}
	\renewcommand{\figurename}{Figure}
	\caption{Cross sections for spin relaxation as functions of electric field at $B=10^{-3}$ T (lower curve),
	$B=0.01$ T (upper curve). $E_\text{coll}=1$~mK for both curves.
	The initial state for $B=0.01$ T is shown by the red (light grey) line in the lower panel of Fig.~1.}
\end{figure}

Figure 5 displays the electric field dependence of spin relaxation cross sections at magnetic fields of $10^{-3}$ and 0.01 T
and collision energy of $10^{-3}$ cm$^{-1}$. The cross sections increase monotonously at low electric fields but exhibit two sudden drops
at $E>0.5$ kV/cm. The location of the dips in Fig. 5 depends on the magnitude of the applied magnetic field. For $B=0.01$ T,
they occur at electric fields of
about 1 and 1.5~kV/cm. These values coincide with the positions of the level crossings marked by the vertical arrows in Fig. 1.
As the electric field increases, the energy of the initial state $|J=\frac{3}{2},M=\frac{3}{2},\epsilon=-1\rangle$ becomes lower
than that of the two $|M|=\half$ states. At electric fields larger than $\sim 2$ kV/cm, both of the $|M|=\frac{1}{2}$ channels become
energetically forbidden, and spin relaxation can only occur via the
$|J=\frac{3}{2},M=\frac{3}{2},\epsilon=-1\rangle \to |J'=\frac{3}{2},M'=-\frac{3}{2},\epsilon'=-1\rangle$ transition.
An analysis of state-resolved cross sections shows that this transition is the least probable of all spin relaxation channels. As a consequence, the total
inelastic cross section decreases by four to seven orders of magnitude depending on the magnetic field.
The data shown in Figs. 2 and 5 demonstrate that spin relaxation of OH molecules in high electric field-seeking states
can be completely suppressed by properly chosen combinations of electric and magnetic fields. The control is especially
robust at low collision energies (on the order of 1 mK) and magnetic fields not exceeding 0.01 T. However, Fig. 1 shows that for any given
value of magnetic field, it should be possible to chose an electric field at which the $|M|=\half$ channels are closed. The necessary electric
field can be determined from the positions of the crossings in the lower panel of Fig. 1.

\section{Summary}

We have developed a rigorous quantum theory for collisions of molecules in $^2\Pi$ electronic states
with structureless atoms in the presence of superimposed electric and magnetic fields. The matrix elements of the
electrostatic potential and molecule-field interactions have been derived in the fully uncoupled representation
of Hund's case (a) basis functions. The theory has been applied to elucidate the mechanisms of inelastic transitions
in low-energy collisions of OH molecules with He atoms. Our results suggest that spin relaxation in collisions
of OH molecules proceeds via direct coupling of different Zeeman states induced by the anisotropy
of the atom-molecule interaction potential. The rate constants for spin relaxation at temperatures above 0.1 K are
on the order of $10^{-12}$ cm$^3$/s, leading to very short trapping lifetimes. We conclude that sympathetic
cooling of OH molecules using cryogenic $^3$He gas at densities $>10^{15}$ cm$^{-3}$ and temperatures
0.1 - 1 K appears unfeasible.

We have demonstrated that spin relaxation of OH molecules at temperatures below 0.01 K can be efficiently
manipulated by electric fields of moderate strength ($\sim$10 kV/cm) available in the laboratory. The mechanism of
control is based on suppressing certain relaxation pathways using superimposed electric and magnetic fields.
This technique can be used to facilitate evaporative cooling of $^2\Pi$ molecules in electromagnetic traps \cite{OHtrap}.
We have found that electric fields modify the collision energy dependence of inelastic cross sections
and may lead to the formation and splitting of Feshbach resonances. The magnetically tunable Feshbach resonances
may be used to create weakly bound He$\cdots$OH complexes. It would be interesting to explore
the effects of magnetic fields on spin-orbit, vibrational, and rotational predissociation of these complexes.
The methods to control inelastic collisions presented in this work may be realized experimentally using
Stark decelerated beams \cite{XeOH,OH1} and electromagnetic traps \cite{Sawyer,OHtrap}.

\begin{acknowledgments}
We thank Hossein Sadeghpour for useful comments.
This work was supported by the Chemical Science, Geoscience, and Bioscience Division of the Office of Basic Energy Science, Office
of Science, U.S. Department of Energy and NSF grants to the Harvard-MIT
Center for Ultracold Atoms and to the Institute for Theoretical Atomic, Molecular,
and Optical Physics at Harvard University and Smithsonian Astrophysical Observatory.
\end{acknowledgments}


\end{document}